\documentclass[letterpapper]{raa}      

\usepackage{graphicx,times, amssymb}    
\begin{document}

   \title{Could GRB170817A be really correlated to a NS-NS merging?}

   \volnopage{Vol.0 (200x) No.0, 000--000}      
   \setcounter{page}{1}          

   \author{D.Fargion
      \inst{1,2}
   \and M. Khlopov
      \inst{3}
   \and P. Oliva
      \inst{2,4,5}
   }

   \institute{
              Physics Department \& INFN Rome1, Rome University 1, P.le A. Moro 2, 00185, Rome, Italy\\
   \and
              MIFP, Via Appia Nuova 31, 00040 Marino (Rome), Italy\\
   \and
             Center for Cosmoparticle Physics Cosmion; National Research Nuclear University MEPHI, Kashirskoe Sh., 31, Moscow 115409, Russia\\
 \and
             Niccol\`{o} Cusano University, Via Don Carlo Gnocchi 3, 00166 Rome, Italy\\
 \and
             Department of Sciences, University Roma Tre, Via Vasca Navale 84, 00146 Rome, Italy
   }

   \date{Received~~2017 month day; accepted~~2017~~month day}

\abstract{The exciting development of gravitational wave (GW) astronomy in the correlation of LIGO and VIRGO detection of GW signals makes possible to expect registration of effects of not only Binary Black Hole (BH) coalescence, but also Binary Neutron Star (NS) merging accompanied by electromagnetic (GRB) signal. Here we consider the possibility that a Neutron Star (NS), merging in a NS-NS or NS-BH system might be (soon) observed in correlation with any LIGO-VIRGO Gravitational Waves
detection. We analyze as an example the recent case of the short GRB 170817A observed by
Fermi and Integral. The associated Optical transient OT source in NGC4993 imply a rare near
source, a consequent averaged large rate of such events (almost) compatible with expected
NS NS merging rate. However the expected beamed GRB (or Short GRB) may be mostly
aligned to a different direction than our. Therefore even soft GRB photons, more spread than
hard ones, might be hardly able to shower to us. Nevertheless a prompt spiraling electron
turbine jet in largest magnetic fields, at the base of the NS-NS collapse, might shine by its
tangential synchrotron radiation in spread way with its skimming photons shining in large
open disk. The consequent solid angle for such soft disk gamma radiation may be large
enough to be nevertheless often observed.
}

   \authorrunning{D. Fargion, M. Khlopov \& P. Oliva}            
   \titlerunning{Could GRB170817A be correlated to the NS-NS merging?}  

   \maketitle

\section{Introduction}           
\label{sect:intro}

On August the 17\textsuperscript{th} 2017 at 12:41:06,47~UT the Fermi Gamma Ray Burst Monitor (GRBM) has been triggered by a weak, short ($\Delta t\sim2$~s) Gamma Ray Burst (GRB) named GRB170817A. The allowed location was found at RA=$176.8^\circ\pm11.6^\circ$, DEC=$-39.8^\circ\pm11.6^\circ$ quite within the Hydra Cluster located at RA$\simeq197.5^\circ$, DEC$\simeq-23.4^\circ$. At nearly the same time (seconds or minutes before) the same Ligo-Virgo (a first triple detection event) had revealed a very peculiar signal possibly associated to a NS-NS merging binary system; the event has been named GW170818. The consequent rush in different ground and space telescopes (Hubble, Chandra among the ones that leaved traces) testify an optical transient pointing toward NGC4993 as the candidate both for the GW and the short GRB170817A \cite{Wikipedia_GRB170817A(2017)}.
The NGC4993 distance is almost 40~Mpc (a near cosmic 0.01 radius).

Most of the astronomical collaborations are keeping the strict silence on their data and records. However even with such a few news and facts we may already address to a key question that possibly is wondering all of us: could GRB170717A be really originated at the same time and by the same source of the (first) NS-NS GW collapse (possibly) discovered by among the earliest Ligo-Virgo triangle array event?

\section{The GRB-GW Connection: an historical discover}
\label{sect:connect}
In the recent two months and few days remarkable events occurred in LIGO-Virgo collaboration: a first published triple signal BH-BH event as GW170814 by \cite{LIGO VIRGO (2017)} and the celebrated Prize Nobel  \cite{Aasi et al.(2015)} \cite{Nobel Physics (2017)} of a few days ago to the pioneer LIGO leaders.It is  the triumph of the proposal of the earliest prophet more than half century ago \cite{Ge62}. However a more radical discover is under the LIGO-VIRGO (August 2017) discovery ashes.
Indeed since last August 2017 the rumors and the few notes about a first Ligo and Virgo triangled event excited all the earliest and late GW hunters, the today known GW170814 event. The second welcomed news (or rumor) was a first NS-NS binary merging signal; it was very exciting and fundamental for  all relativity theorists. However, the third, almost unexpected, news of a short $\gamma$ event GRB170817A, possibly quite correlated in space sky and in time with the prompt GWs has been embraced, by some of us, with enthusiasm but also with many worries and reservation. Indeed this GW-GRB connection would be the goal in the wild dreams of high energy astrophysics community.  The NS-NS (as well as a more rare NS-BH) collapse may lead to a baryon (and also a lepton) load explosion  whose signal, for most of the GRB model, would lead to a beamed gamma jet. Therefore, most of the authors (including us \cite{Fargion:2016}) were quite skeptic to forecast a soon (near years) eventual NS-NS correlated GW (almost spherical) with a correlated (quite beamed) GRB events. Indeed, we imagined that many years or even decades would be needed for NS-NS GWs detection before one of those shows a lucky beamed jet illuminating toward us at once triggering our $\gamma$ satellites. In an historic prospect we remind the $1967-1997$ long time period lost from the earliest GRB discover via several military satellite triangulation up to the very late and tuned  Beppo-Sax discover of a narrow X ray afterglow : its traces were leading to a consequent correlation with an optical transient and finally to a guest galaxy identification. The probability for such coincidence for a twin or  a triple GW arrays detection, we argued, it is in a first approximation just the solid angle of the jet beam divided by the whole 4$\pi$ of the sky. Because most of the GRB models require an unique $\Delta\Omega/4\pi\approx10^{-3}\div10^{-6}$ shot beamed  (fountain  \cite{1986ApJ...308L..43P} or collimated) Fireball \cite{1997ApJ...485L...5W} or a much thinner, but persistent and thin precessing jet $\Delta\Omega/4\pi\approx10^{-8}-10^{-7}$ \cite{1999A&AS..138..507F} \cite{Fargion:2016}, the appearance of a visible GRB in the first NS-NS gravitationally detected event seemed very remote \cite{Fargion:2016c}. Moreover, the GRB 170817A and its possible correlated GW NS NS merging event exhibit several extreme  values all at once:
\begin{enumerate}
\item[a)] it is the nearest short GRB (SGRB) ever detected (40 Mpc);
\item[b)] it is among the weakest fluency SGRB ever recorded;
\item[c)] it is the most soft SGRB in whole known SGRB catalogues.
\item[d)] it is the first NS NS (or more rare NS BH) merging GW event ever detected
\item[e)] it is the nearest GW event observed
\item[f)] It is  the first NS-NS merging into a new born NS or  a new BH.
\item[g)] It is (probably) the most powerful energy fluency  GW signal ever observed by LIGO-Virgo
\item[h)] It is the very first GWs correlated with an electromagnetic burst and afterglows ($\gamma$, X, optical)
\end{enumerate}
Therefore following all these exceptional rare features, in particular the last one, a main question rises: could GRB170817A really be correlated to GW170818?
In the following, even without the huge amount of ( we imagine, yet unpublished) data, we shall enumerate and summarize in more details the main reasons that tempted us, at first glance, not to accept such a correlation. We shall however also show  that these numerous and extreme behaviors of GRB170817A make it surprisingly a possible (some how lucky) realistic candidate for GW-GRB correlation, assuming a natural extension of the new born synchrotron radiation jet in NS NS explosive expansion: its  radiative solid angle beam, at earliest stages, it is not constrained in a narrow beam jet, but it is much widely spread, because of the ring spirals of the electrons,  into a twin open cones or gamma disks. At lowest energy these disks are almost enlarged,  offering a more large spread signal and a higher probability to hit us off axis, almost with no much amplified in their apparent output. The late relativistic jet electrons in advanced spirals rings, more collinear with the jet axis itself,  are shining, by synchrotron or Inverse Compton Scattering, into a twin  narrow cones able to blaze us into a much rare solid angle, mostly from the vast cosmic volumes where there are many more sources. The jet trembling by its spin and its precessing processes  during the NS-NS accretion and collapse, it leaves these imprints in the fast  multi-peak variability of GRB, its possible X flash precursor and its late re-brightening along late GRB afterglows \cite{2001foap.conf..347F},\cite{Fargion:2006we}.

\section{The Exceptional GRB170817A signatures}
\label{fig:integral}

\subsection{The GRB Fluence}
\begin{figure}[t!]
\begin{minipage}[t]{0.495\linewidth}
 \centering
\includegraphics[width=53mm,height=32mm]{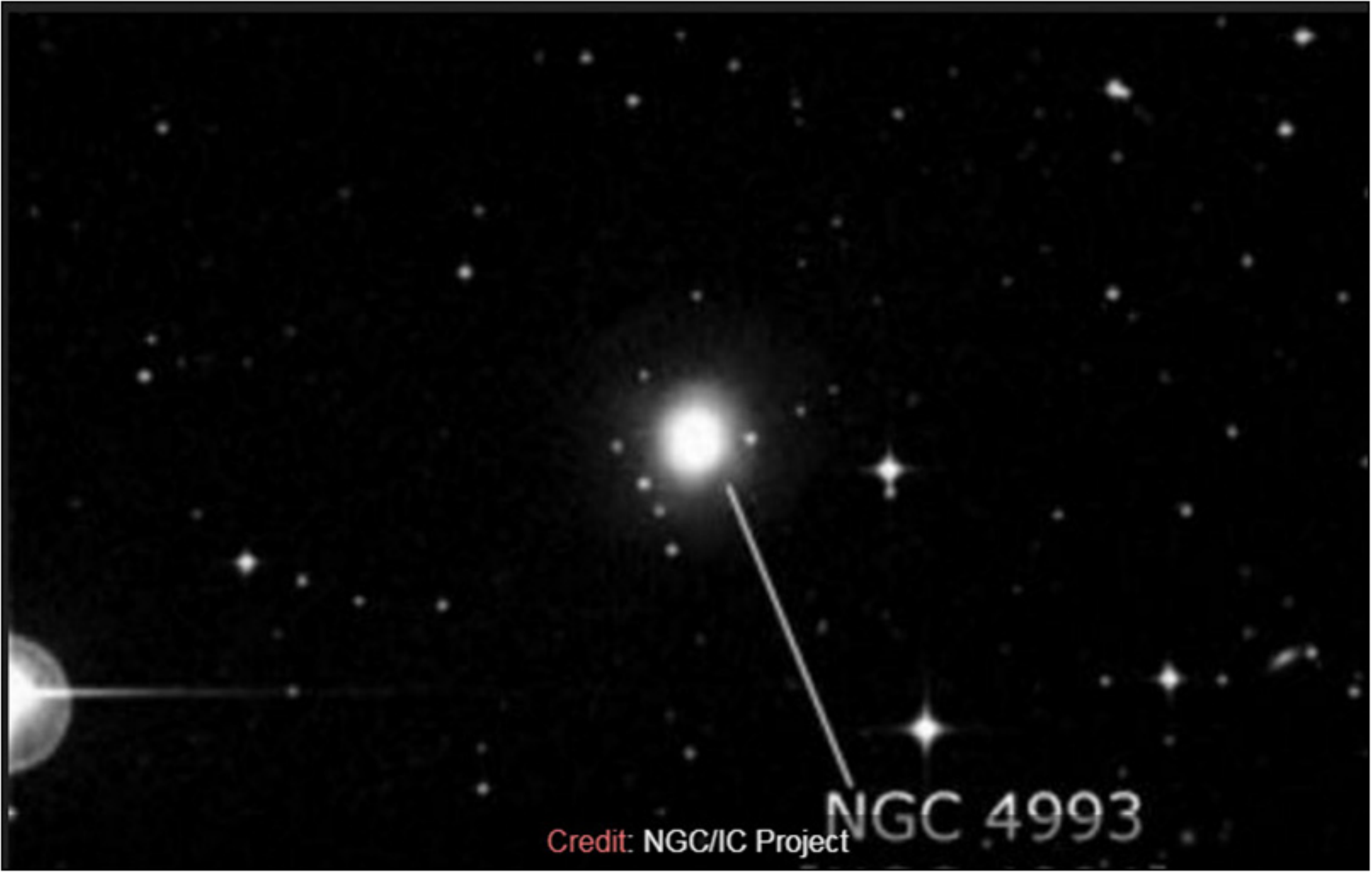}
 \caption{ The Galaxy NGC4993, at 40 Mpc distance from us, where the GRB 170817A had been localized }
 \label{Fig1a}
\end{minipage}%
\begin{minipage}[t]{0.495\textwidth} \centering
\includegraphics[width=53mm,height=52mm]{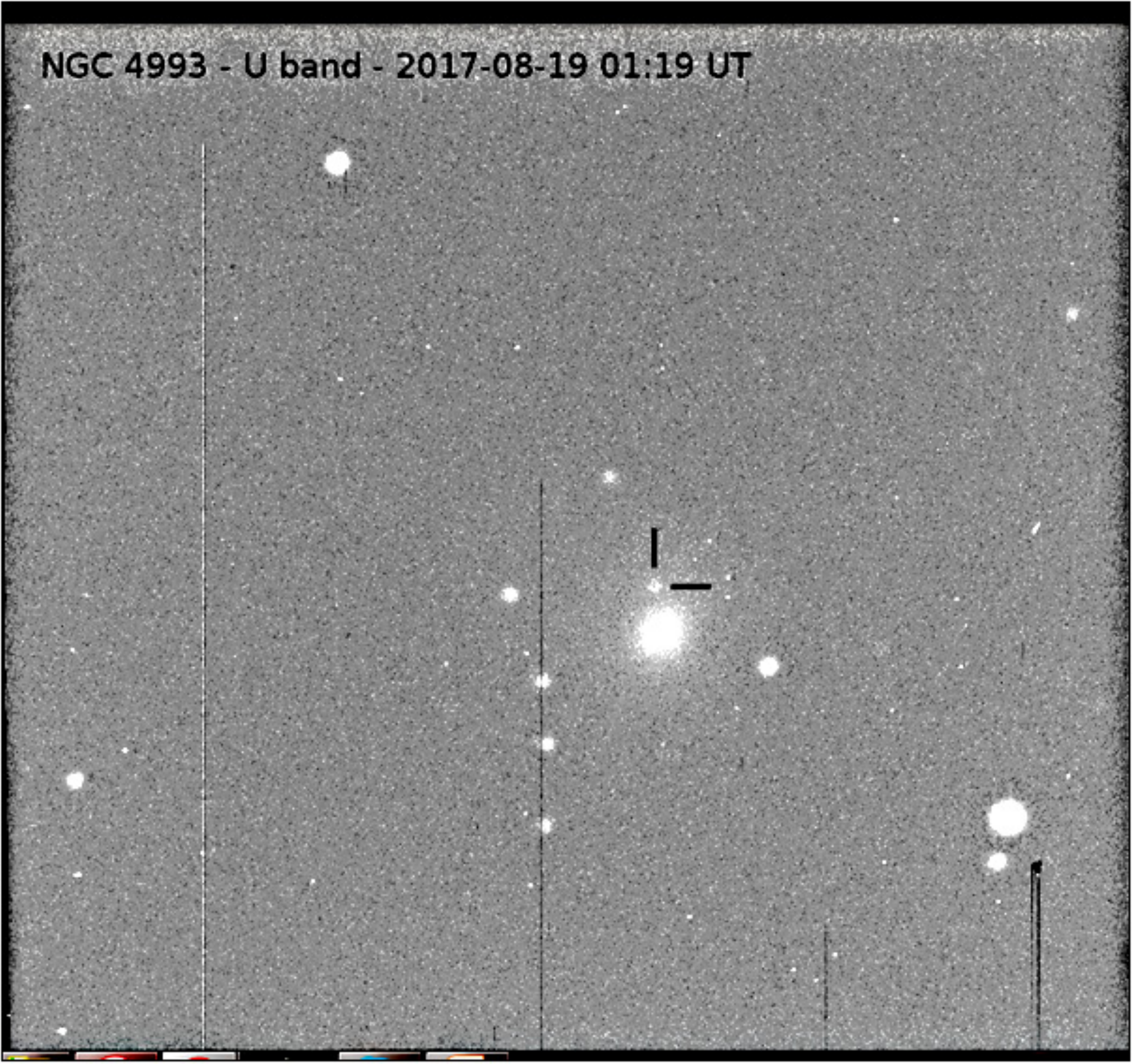}
\caption{The probable OT of the GRB 170817A observed the day after, on 19 August 2017, at Galaxy NGC4993.
 The exact X ray afterglow and the Optical Transient (OT) evolutions are still unpublished.}
 \label{Fig1b}
\end{minipage}%
\end{figure}

\begin{figure}[t!]
 \begin{minipage}[t]{0.495\textwidth}
 \centering
  \includegraphics[width=53mm,height=52mm]{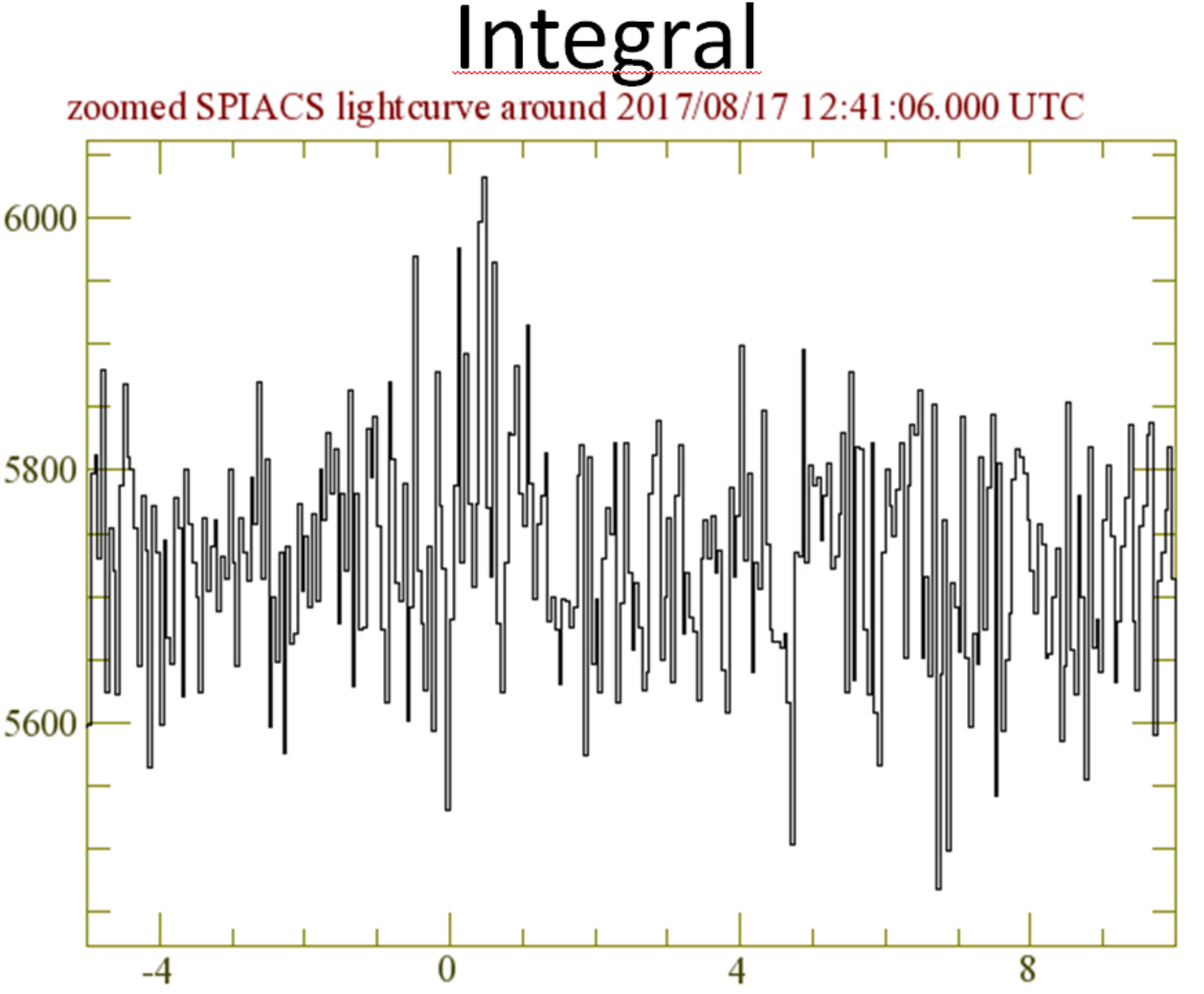}
 \caption{The Integral gamma burst detection signal in correlation with the Fermi GRB 170817A.
   Note the great noise over the  signal, whose exact statistical weight it is not yet published.}
 \label{Fig1c}
\end{minipage}%
  \begin{minipage}[t]{0.495\textwidth}
  \centering
   \includegraphics[width=53mm,height=52mm]{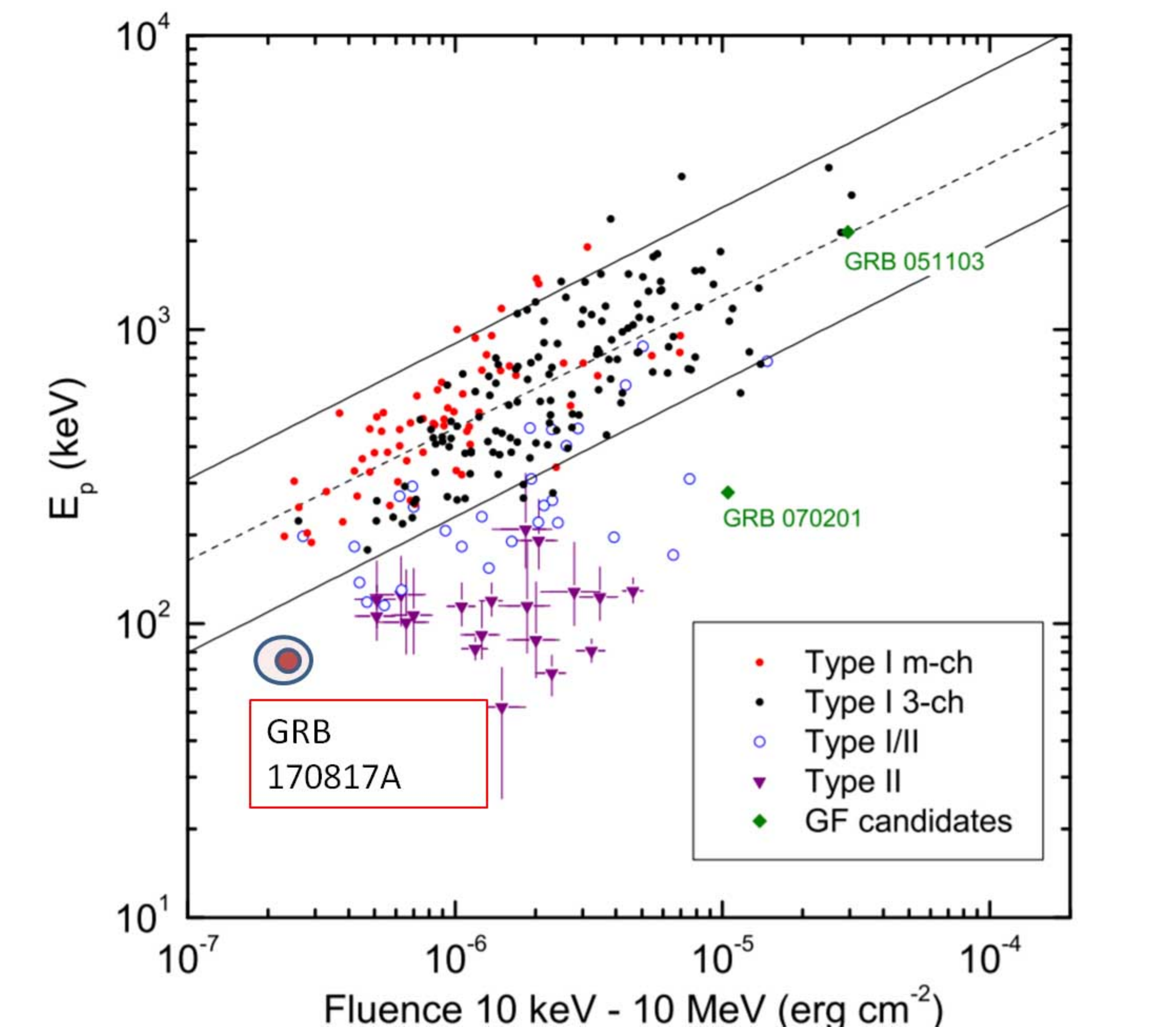}
  \caption{The extreme low Fermi GRB170817A energy fluency for its observed peak energy respect to all the previous ones}
\label{Fig1d}
  \end{minipage}%
\end{figure}

The extreme behavior of the GRB 170817A may be seen at once while comparing its energy fluency respect to its observed peak energy. The unique event record of GRB 170817A that we may show it is the Integral one : See Fig. \ref{Fig1c}.
Considering the candidate source NGC4993 , see Fig. \ref{Fig1a}, \ref{Fig1b}, at a  distance ($40~Mpc$), the Fermi energy fluency $\phi_{\textrm{Fermi}}=(2.3\pm0.4)\cdot10^{-7}$ $~erg~cm\textsuperscript{-2}$, it does correspond to an isotropic output energy
\begin{equation}
E^{\textrm{iso}}_{\textrm{GRB170817A}}=(4.14\pm0.6)\cdot10^{46}\,\textrm{erg}
\end{equation}

\begin{figure}[h]
  \begin{minipage}[t]{0.495\linewidth}
  \centering
   \includegraphics[width=59mm,height=52mm]{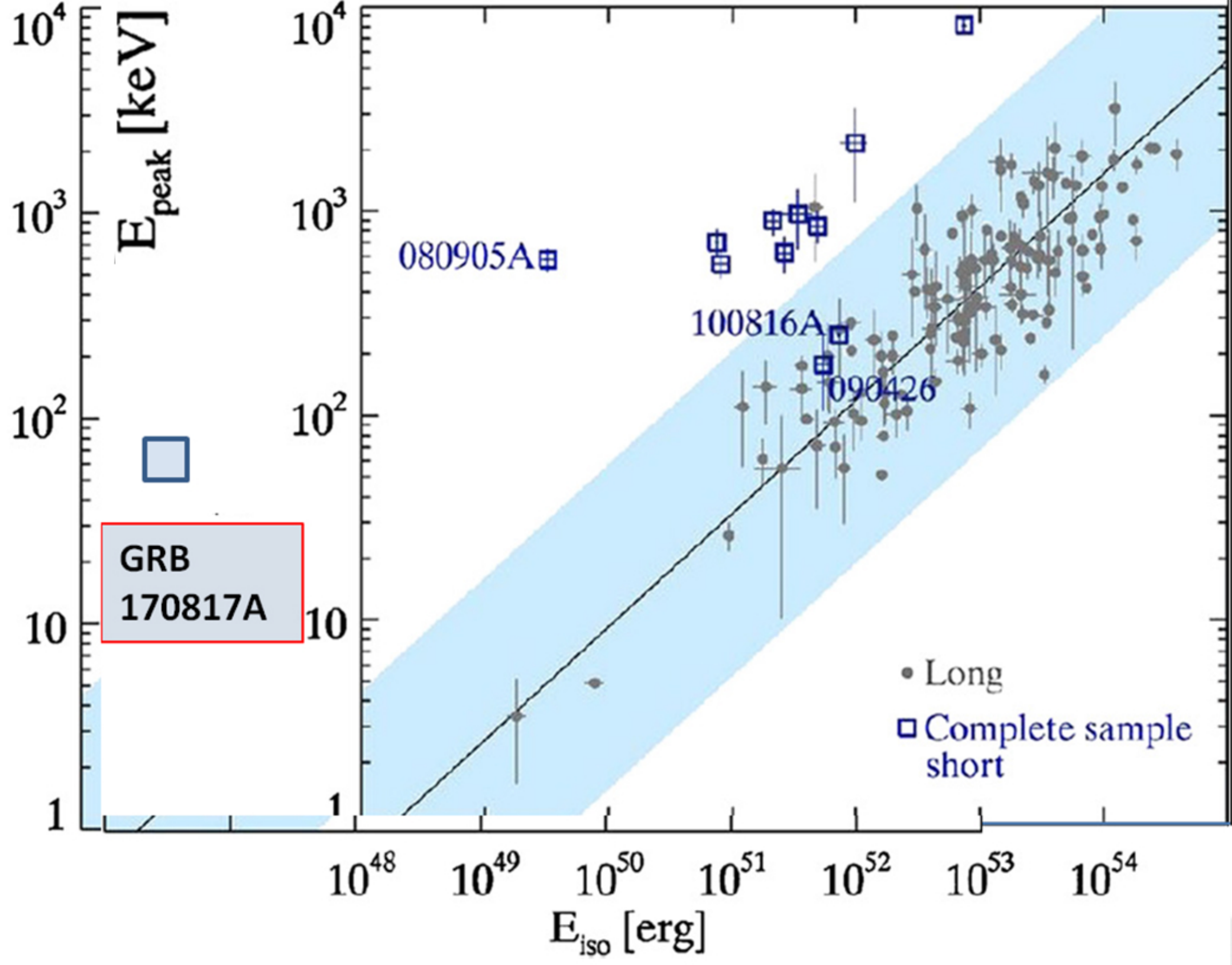}
   \caption{The extreme low GRB 170817A isotropic Energy respect the up to day known ones }
  \end{minipage}%
  \label{Fig:ghirlanda1}
  \begin{minipage}[t]{0.495\textwidth}
  \centering
   \includegraphics[width=53mm,height=52mm]{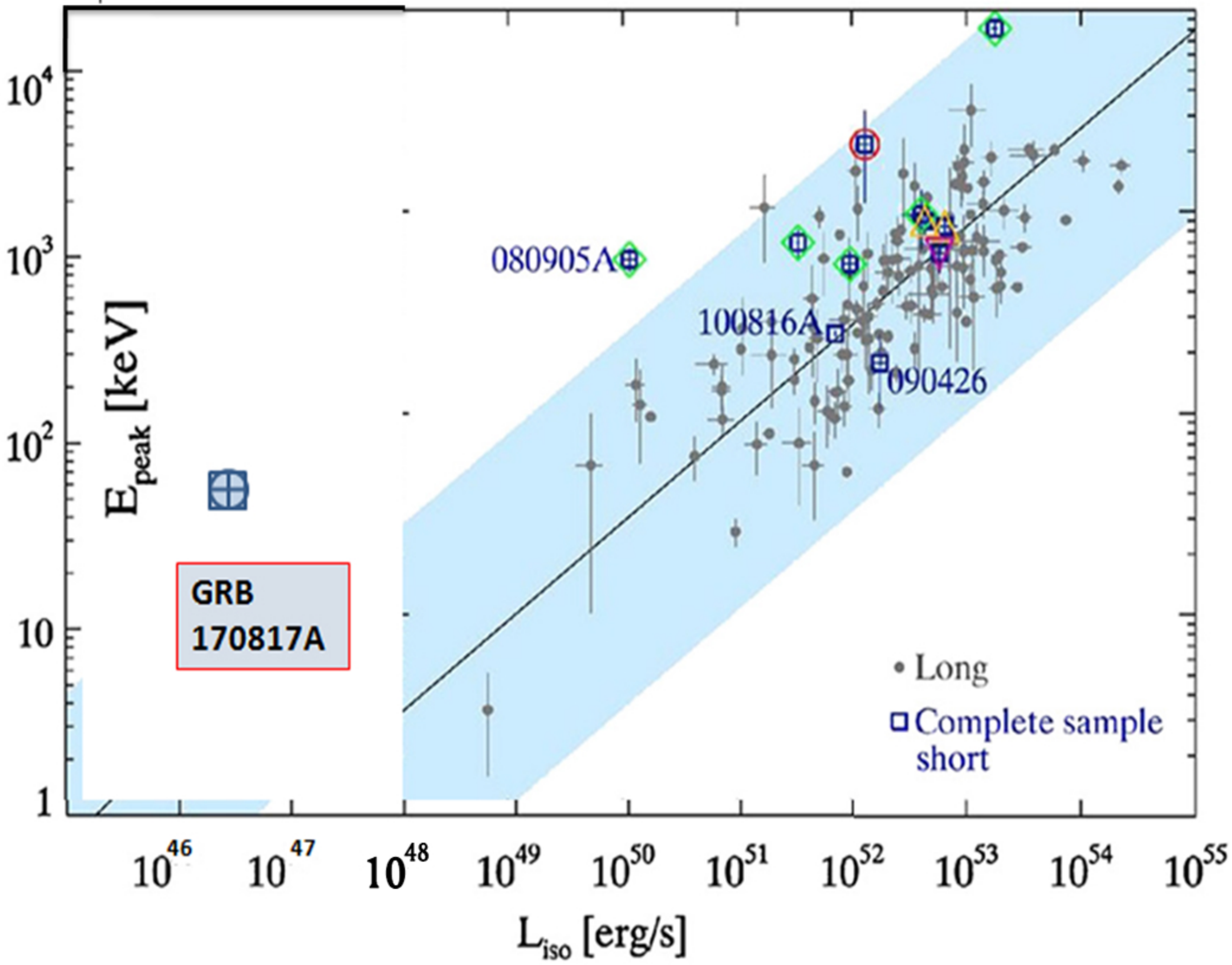}
  \caption{The extreme low GRB 170817A isotropic Luminosity respect the up to day known ones }
  \end{minipage}%
  \label{Fig:ghirlanda2}
\end{figure}

A first estimate based on the  GRBM area detector and on the error width imply that Fermi \cite{FERMI GNC: 2017} observed several hundreds of photons. The additional counts by Integral Satellite  contains several thousands of events with a signal/noise, we estimated, of about $3\div4$ sigma.
Because of the short GRB duration the GRB isotropic luminosity peak is nearly
\begin{equation}
L^{\textrm{iso}}_{\textrm{GRB170817A}}=(2.07\pm0.3)\cdot10^{46}\,\textrm{erg}\,\textrm{s}^{-1}
\end{equation}
The signals (at average peak $E_p\simeq 82$~keV) may be compared with most updated catalog by \cite{G.Ghirlanda et al.(2015)} and  \cite{Svinkin et al(2016)} for short GRBs up to 2016, as shown in figure Fig. \ref{Fig:ghirlanda1}, \ref{Fig:ghirlanda2}

\begin{figure}[h]
  \centering
   \includegraphics[width=133mm,height=82mm]{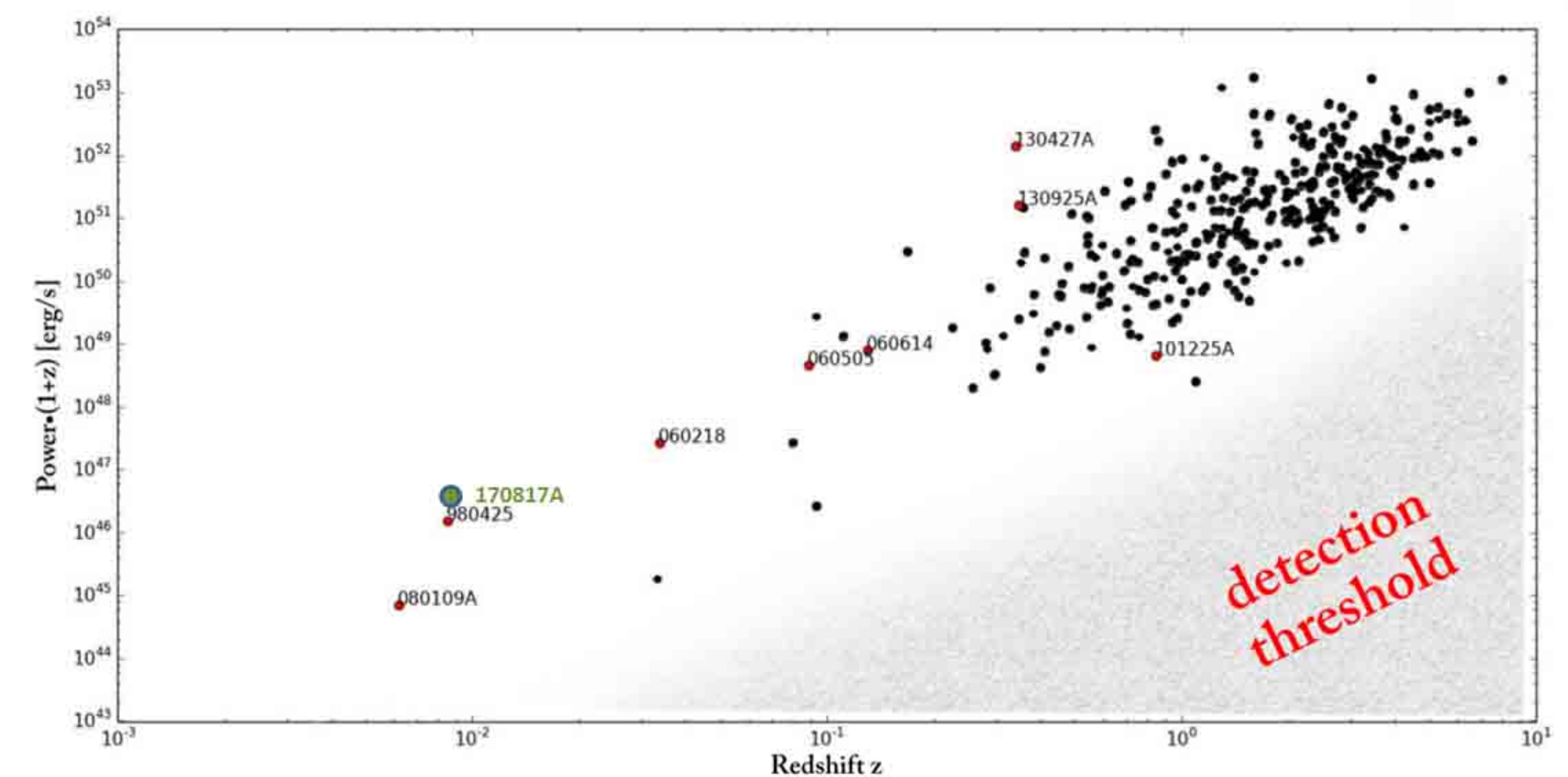}
   \caption{The extreme near distance for the extreme low luminosity of GRB 170817A respect to most known ones.
   Note the rare nearest and almost overlapping low luminosity of both GRB 25 April 1998 GRB and the  on 17 August 2017 one.
   As for the earliest  event \cite{1999A&AS..138..507F} we propose once again an off axis nature of these detections}
  \label{Fig:ghirlanda}
\end{figure}

\begin{figure}[h]
  \begin{minipage}[t]{0.495\linewidth}
  \centering
   \includegraphics[width=73mm,height=52mm]{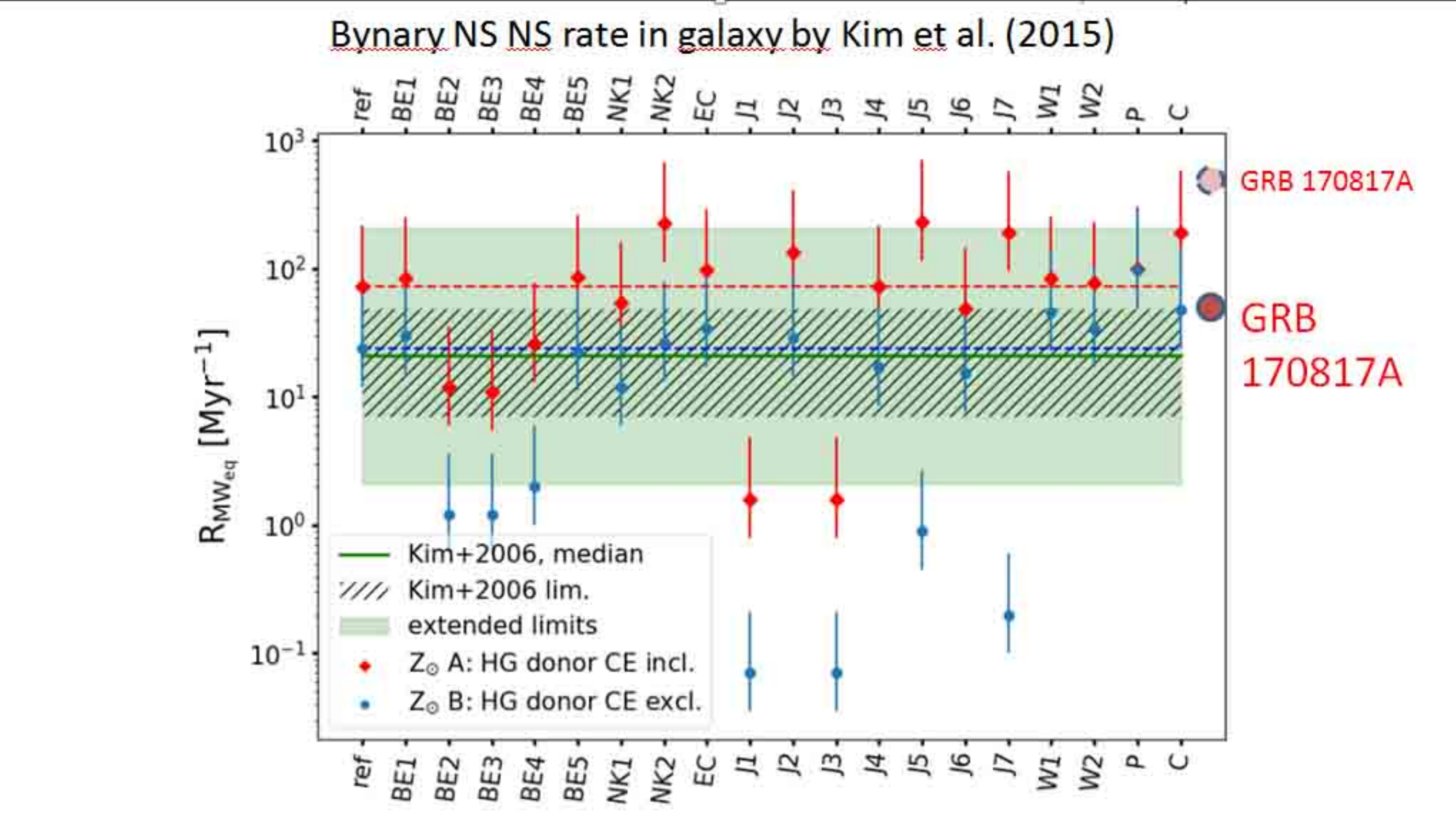}
   \caption{The consequent NS-NS collapse rate derived by the unique GRB170817A event, assumed one a year rate respect to the earlier models. The dashed dot describe the case of a galaxy number density of 0.1 for each Mpc cube. The solid dot assume one galaxy within each Mpc volume.}
  \end{minipage}%
  \label{Fig:ghirlanda5}
  \begin{minipage}[t]{0.495\linewidth}
  \centering
   \includegraphics[width=73mm,height=52mm]{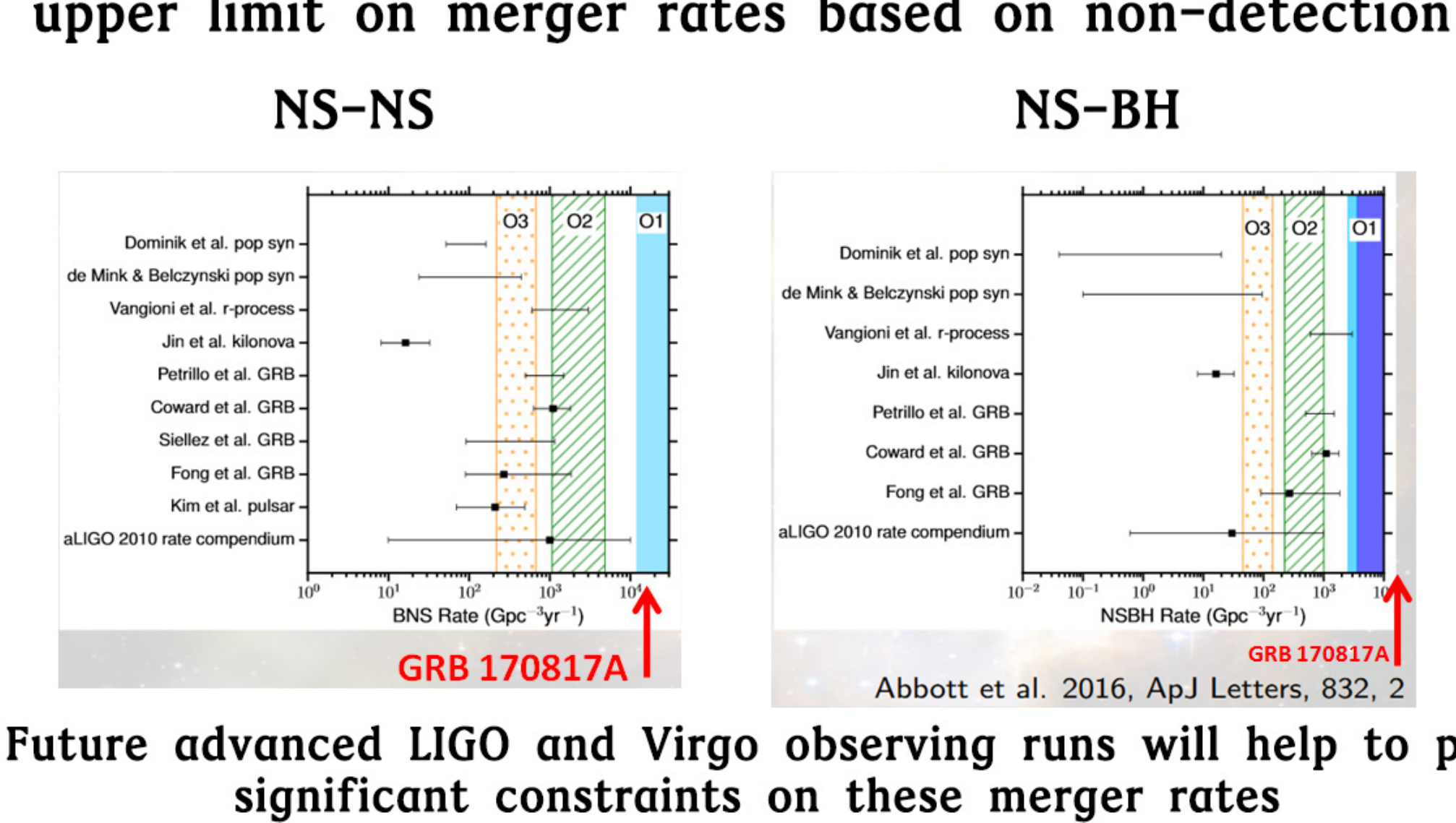}
   \caption{The consequent NS-NS collapse rate derived by the unique GRB170817A event, assumed as in earlier figure, one event each year, respect to  earlier rate models derived by chemical pollution of NS NS via r-processes in our galaxy }
  \end{minipage}%
  \label{Fig:ghirlanda6}
\end{figure}

\subsection{The very nearby distance and weakest power of GRB~170817A}

A surprising feature of the event it is its nearest location respect all the observed ones
and its weakest output . Indeed, the following peak luminosity versus the red-shift diagram shows the rare near location of the GRB 170817A and its lowest luminosity.

A possible explanation for the ability to discover such a near (and weak) GRB may be the knowledge of the GW17018 whose narrow timing and thin correlated directionality had forced Fermi team to enhance and remark a signal possibly otherwise lost as a noise.

\subsection{The correlation in space and time of GRB and GW}

Indeed, the strength of the Fermi discover (and Integral) might be the narrow time correlation (tens seconds?) and the narrow sky location (tens of hundred square degree). However, these details are still unknown: we can only imagine that the need to kidnap the main whole telescope of the world to the NGC4993 had to be very compelling. Therefore, the a priori probability of the GRB-GW connection is hidden in the exact time-solid angle correlation and their error box, is very probable , even it is  still unpublished.

However, we shall continue to analyze the self consistence of a such (premature or exceptional) discover. In particular we now wonder on the possible consistence of such a nearby (40~Mpc) NS-NS event within, let's say, about one years of LIGO-Virgo lifetime.

\subsection{The NS-NS cosmological expected rate versus the nearby NGC4993 observed event}

The NS-NS collision and/or merging rate (as well as a more rare NS-BH - of a few solar masses) it has been a wide subject of study in the last decades. It combines the r-processes chemical pollution of heavy nuclei in our galaxy, the known binary NS-NS system and their lifetime, the kick of NS or BH after the merging.
Because of the present (unique) GRB-GW event and its distance  we may also offer a very rough  estimate of this rate by such nearby volume. The LIGO-Virgo recording (we assume 1 year even if Virgo was active only since and within August 2017), the 40 Mpc distance imply an average NS-NS collapse as large as

\begin{equation}
R_{\mathrm{NS-NS}}\simeq1.56\cdot10^4\,\mathrm{Gpc}^{-3}/\mathrm{yr}
\end{equation}
This value is quite large (respect to most model expectation). The same result may be re-scaled for each galaxy assuming a given average cosmic (or local) galaxy density in each cube Mpc: assuming as usual 0.1 galaxy per Mpc, the rate becomes

 \begin{equation}
R_{G_{\mathrm{NS-NS}}}\simeq160\,\mathrm{Myr}^{-1},
\end{equation}
 while assuming a galaxy per Mpc density, we derive a much lower and a more acceptable one:
\begin{equation}
\bar{R}_{G_{\mathrm{NS-NS}}}\simeq 16\,Myr^{-1}
\end{equation}
Let us remind that following \cite{2004ApJ...612..364N}, the galactic density may be $10^{-2}$ Mpc $\textsuperscript{-3}$, leading to an exceptional $R_{\mathrm{NS-NS}}\simeq1600$ $~Myr\textsuperscript{-1}$. This last rate,if taken as the real value, it makes very questionable the GRB-GW connection. Let us show these values within recent model predictions.

\begin{figure}[h]
\begin{minipage}[t]{0.495\linewidth}
  \centering
   \includegraphics[width=140mm,height=90mm]{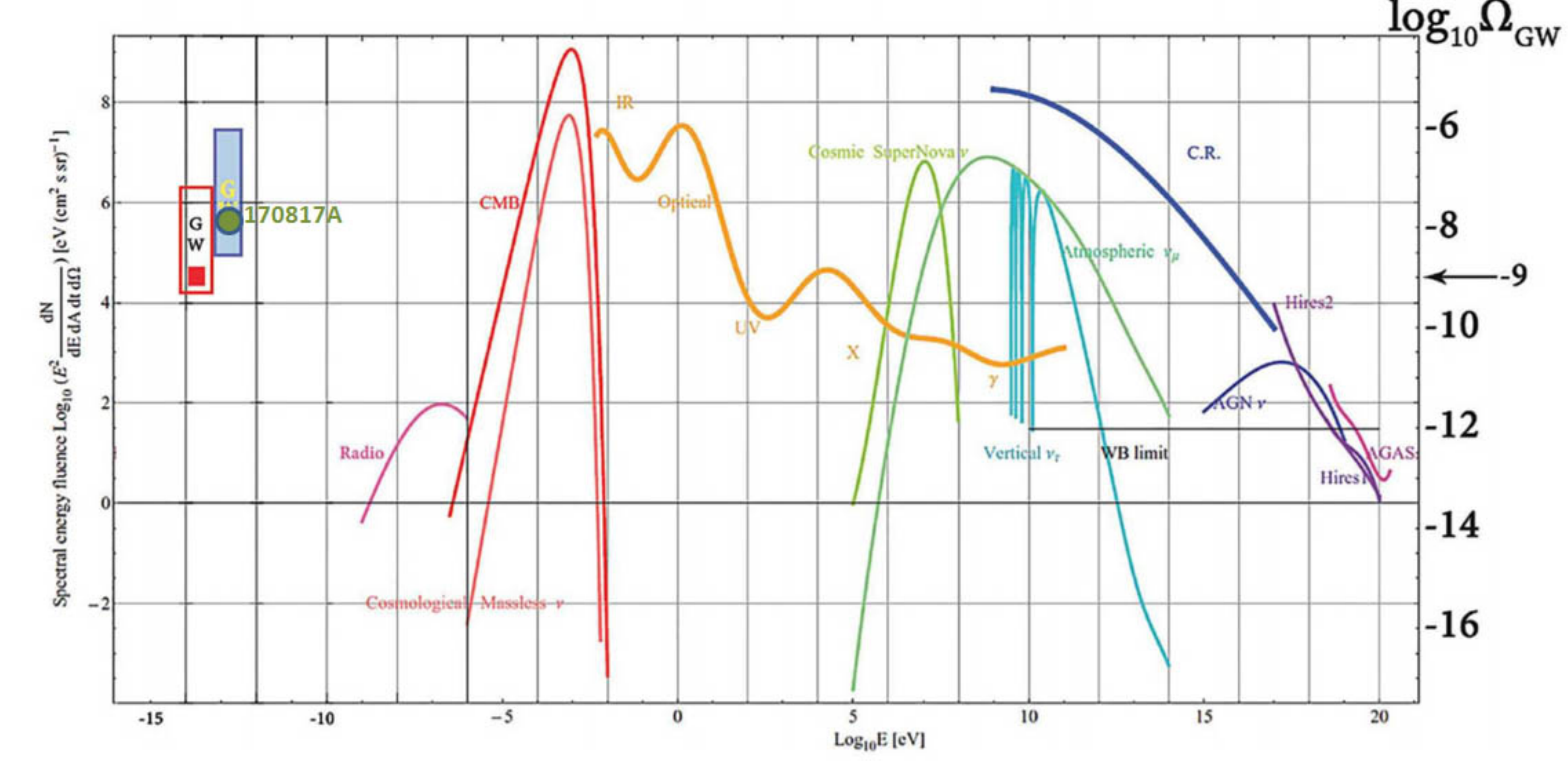}
     \end{minipage}%
      \caption{The GW GRB170817A energy density contribute by extrapolating its $ (40 Mpc)^{3}$ volume and an event a year on average along all the universe.}
  \label{Fig:fluency}
  \end{figure}

\begin{figure}[h]
  \begin{minipage}[t]{0.495\linewidth}
  \centering
   \includegraphics[width=140mm,height=90mm]{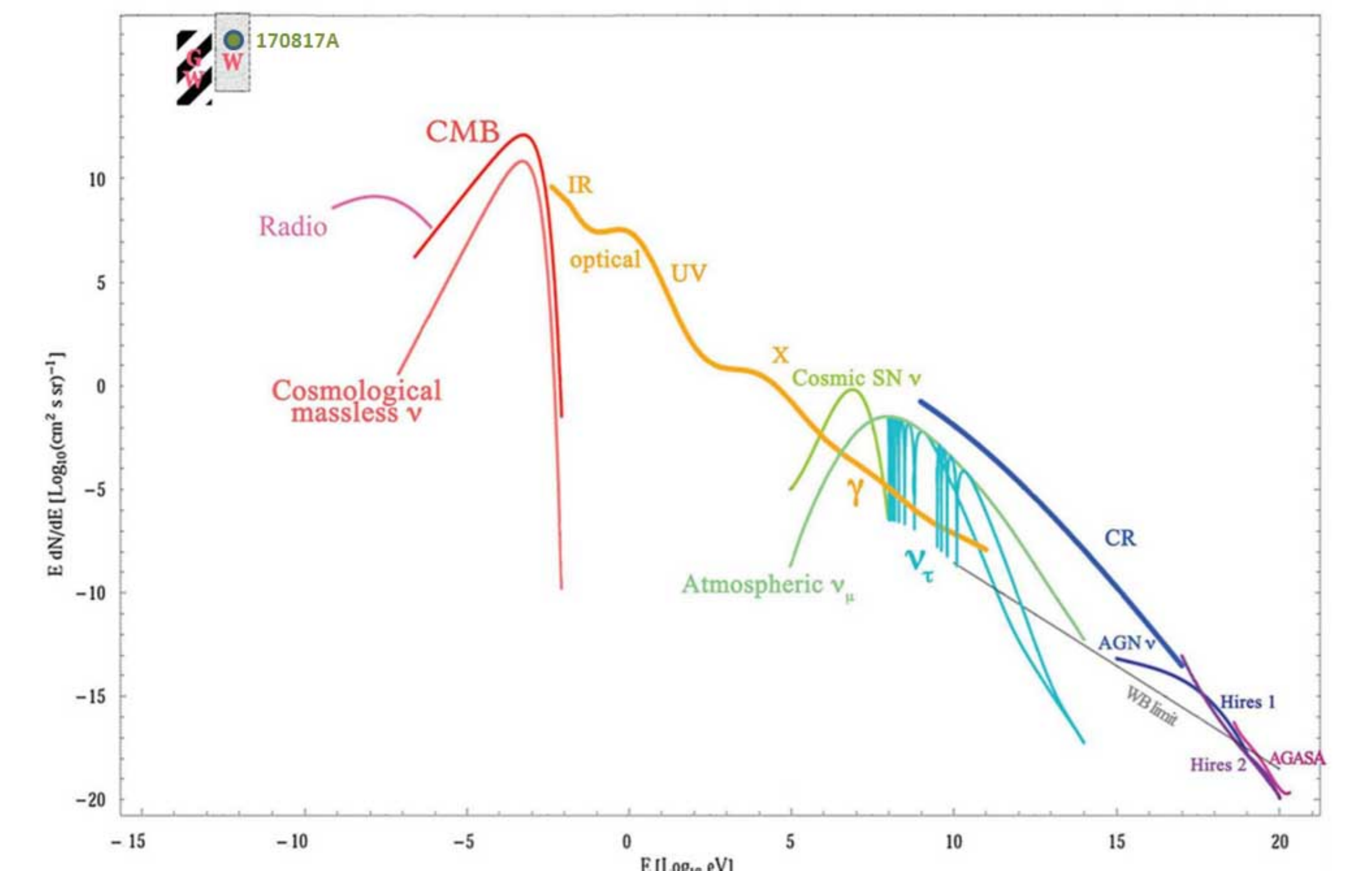}
    \end{minipage}%
       \caption{The GW GRB170817A flux number density contribute by extrapolating its $ (40 Mpc)^{3}$ volume and an event a year on average along all the universe.}
  \label{Fig:flux}
\end{figure}

As a curiosity, one would like also to encompass the whole average energy fluency (or the corresponding cosmic energy density) indebt to this averaged NS-NS event (see Fig. \ref{Fig:fluency}) with known photon and particle background. The same may be done with the averaged flux number made by such cosmic signals  (see Fig. \ref{Fig:flux})

To make this estimation on figure~\ref{Fig:ghirlanda} one should know the GW energy density signal (which has not been yet released). But it is easy and possible to offer an approximated estimation  assuming that two NS-NS collapsed into a new one of 2~M$_\odot$ (once a year at 40 Mpc). The gravitational binding energy of NS-NS merging nearly comparable to the NS binding energy one   \cite{LANDAU1975345}. More precisely, assuming one initial solar mass, with initial radius of 10 km, assuming the same nuclear NS mass density before and after the merging; note that $R_{\mathrm{NS}}^{\mathrm{final}}\simeq\sqrt[3]{2}\cdot10$ km:
\begin{equation}
\Delta E_{2\,\mathrm{M}_{\odot}^{\mathrm{final}}}\simeq\frac{3}{5}\frac{G(2\mathrm{M}_\odot)^2}{R_{\mathrm{final}}}-2\Delta E^{NS}_{1\,\mathrm{M}_\odot}
\end{equation}
Taking into account that $\Delta E^{NS}_{1\,\mathrm{M}_\odot}\simeq2.6\cdot10^{53}$ erg, the result becomes
$$
\Delta E_{2\,\mathrm{M}_\odot}\simeq\left[\frac{4}{\sqrt[3]{2}-2}\right]\Delta E_{\mathrm{NS}}\simeq1.174\Delta E_{\mathrm{NS}}\left(\frac{M}{M_\odot}\right)^2\simeq3\cdot 10^{53}\,\mathrm{erg}\left(\frac{M_f}{2M_\odot}\right)^2
$$
The expression above scale for a final 3 solar masses of final NS mass (one of the maximal allowed NS mass) into
\begin{equation}
\Delta E_{\mathrm{NS}}\simeq 6.8\cdot10^{53}\,\mathrm{erg}\left(\frac{M_{NSf}}{3M_\odot}\right)^2
\end{equation}

There are good argument to believe that the needed energy it is this or a bigger one. For sake of simplicity we assumed both  initial neutron star are about 1.4 solar mass  and their final mass it is near $M_{M_f} = 2.7 M_\odot$; the energy out put is:
$ 5.4\cdot 10^{53}\,\mathrm{erg}\left(\frac{M_f}{2.7M_\odot}\right)^2$ leading to an expected $\Phi_{\mathrm{GW170818}}\simeq 2.8~erg~cm\textsuperscript{-2}$ on Earth, one of the largest energy fluency ever observed. However depending on the amplitude of the signal it may be that the GRB170817A GW was making a BH. This average fluency value may be scaled from 40 Mpc to 4 Gpc after the whole average $4\pi$ again assuming such 1 event each year in each 40 Mpc volume. One thus obtains approximately  for the whole universe (see Fig. \ref{Fig:fluency}):
$\Phi_{\mathrm{GW}}^{\mathrm{cosmic}}\rangle=5\cdot10^5$ $~eV~cm\textsuperscript{-2}~s\textsuperscript{-1}~sr\textsuperscript{-1}.$
One may also remind that the peak frequency of NS-NS GW  emission are at near kHz ($\sim 4\cdot 10^{-12}$ eV) frequency and consequently one derives the number of NS-NS graviton rate reaching us from the whole universe, getting a huge flux number
\begin{equation}
\langle\dot{N}_{\mathrm{GW}}^{\mathrm{NS}}\rangle\simeq1.2\cdot10^{17}\,\mathrm{cm}^{-2}\mathrm{s}^{-1}\mathrm{sr}^{-1}
\end{equation}
This number density (see Fig. \ref{Fig:flux}) exceed by 5 orders of magnitude the Cosmic Black Body photon rate, so that we are hit by much more cosmic gravitons than cosmic photons. Such a flux  number rate it is comparable only with our (very local) solar sunny photon rate.

\subsection{The gravitational radiation NS-NS scale and luminosity}
We remind that all the details  of the LIGO-Virgo observations are not yet available for us. Therefore we present here a simplest estimate of the NS-NS luminosity to fix also a consistent time scale of the final collapse:
\begin{equation}
L_{\mathrm{binary}}\simeq3.5\cdot10^{53}\,\mathrm{erg}\mathrm{s}^{-1}\left(\frac{\mathrm{M}}{\mathrm{M}_\odot}\right)^{10/3}\left(\frac{3\,\mathrm{ms}}{P}\right)^{10/3}
\end{equation}
From here we imagine that most of the NS-NS emission took place in 3-10 ms scale time. Very possibly a late train of GW was very long and precise. However most numerical GW simulation ignore the binary coalescence feeding an accretion disk and a main jet companion birth and activity.

The exact inner jet structure and its output (luminosity) in time and its opening solid angle it is still very controversial. We did discuss and defend since a long time \cite{1999A&AS..138..507F} the inner precessing thin jet whose opening angle may justify the wide spectrum of all the apparent GRB luminosity, a range of nearly eight orders of magnitude, for the example, from present $L_{\mathrm{iso}}^{170817A}\sim2\cdot10^{46}$~erg s\textsuperscript{-1} to the most powerful and usually cosmic ones $L_{\mathrm{iso}}\simeq2\cdot10^{53}\div2\cdot10^{54}$ $~erg~s\textsuperscript{-1}$.
However to justify the near and rare GRB on 25 April 1998 at 40 Mpc distance, an off axis blazing has been considered \cite{1999A&AS..138..507F}.

\subsection{The jet dressing in space and time during a GRB}

The observed brightness of GRB, both short or long, shows an (apparent) spread of the luminosity values by variability intensity isotropic output, often exceeding the already enormous SN one by an order of magnitude. Moreover, the wide differences between
GRB and GRB are leading to several puzzles about the possibility to unify all these (apparent) super explosion in an unique model.

We remind that there are at least two kind of GRB: the short ($<2$ s) and the long ones ($>2$ s). We considered as the main solution for all GRB this thin precessing jet whose starting output is due to a NS-NS (for short ones) or (NS-BH) merging collapse (for long ones) \cite{Fargion:2016}.
The requirement \cite{1999A&AS..138..507F} that the GRB  should shine in a narrow collimated ($\frac{\Delta\Omega}{\Omega}\sim10^{-6}\div10^{-8}$), spinning ad precessing jet it is mainly based on the simple assumption that all the GRBs are ejecting and consuming the same $\sim$ NS mass, but the apparent $L_{\mathrm{iso}}^{\mathrm{GRB}}\simeq10^{46}\div10^{54}$ $~erg~s\textsuperscript{-1}$ luminosity peak might be accommodated mostly by a narrow beam whose angular size should be at highest energy
$$
\left(\frac{\Delta\theta}{4\pi}\right)\sim\left(\frac{\Delta\Omega}{\Omega}\right)^{1/2}\sim10^{-3}\div10^{-4}.
$$

The beaming, spinning and (multi) precessing jet, guarantee the fast dynamical variability of the GRBs (a time as short as $\lesssim$ ms), the multi peak trembling output,  while the persistence and the slow decay in time would explain the longest afterglow activities. The special relativity guarantee that the same processes that produce a hard X ray or a MeV-GeVs $\gamma$  photon it is produced within a narrow cone: given an electron (and/or a pair) relativistic jet at Lorentz factor $\gamma_e$ we expect that the out-coming gamma jet angle it is confined into $\Delta\theta\sim\gamma_e^{-1}$. Indeed, we know that following the relativistic Inverse Compton Scattering  (ICS) \cite{Fargion 1998} or the Synchrotron Radiation (SR) \cite{Ginzburg:1965} the opening angle obey at relativistic regime to a similar law: $\Delta\theta\sim\gamma_e^{-1}$, the consequent apparent Luminosity increases as $L_{ph}\simeq\gamma_e^2L_{\gamma_0}$. (The Luminosity power $L = \frac{dE}{dt}$ it is a relativistic invariant).

The final peak photon spectra is both in ICS or SR peaked at $E_{\gamma}^{\mathrm{MAX}}\simeq\gamma_e^2kT$ (assuming a thermal target photon in ICS) or
$E_{\gamma}^{\mathrm{MAX}}\simeq\frac{3}{2}\frac{\gamma^2eB}{m_ec}$ in the synchrotron radiation case. In the latter case
the target photon energy density are due to the  magnetic field $\simeq B^{2}$ one. This well knows similitude ICS $\leftrightarrow$ SR may be understood considering the SR as the ICS of relativistic electron onto virtual photon (by the magnetic field $B$) at Larmor (or Lienard) frequency. Nevertheless, the main result once again leads to $\Delta\theta\sim\gamma_e^{-1}$, $L_{\gamma}^{\mathrm{MAX}}\simeq\gamma_e^2E_{\gamma_0}$.

The average observed power dE/dt is a relativistic invariant,
therefore the observed blazing one is enhanced by the narrow opening jet solid  angle, for example

$\left(\frac{\Delta\Omega}{4\pi}\right)=\left(\frac{\Delta\theta^2}{4}\right)
\sim\gamma_e^2\simeq10^{-7}\left(\frac{E_e}{1.5\,\mathrm{GeV}}\right)^2$; or $\simeq10^{-3}\left(\frac{E_e}{15\,\mathrm{MeV}}\right)^2$.

Therefore the softer and less collimated (and less amplified) photons are more easily observed off axis by $\gamma$ or X detecors.
Naturally a jet produced by the spiral collapse of a NS-NS or a NS-BH is not built by a mono-energetic electron beam, but by  power law one. The most probable electron jet number output might be $\frac{\mathrm{d}N_e}{\mathrm{d}E_e\mathrm{d}t}\propto E^{-\alpha}$ (where $\alpha$ index is usually considered around 2, as the Fermi one), up to a maximal electron energy, about several or a hundred GeVs ones. The maximal energy of the jet electron beam should be able to explain at the same time the GRBs wide spread hardness (from keV to GeVs energies) and the wide apparent luminosity due to the same source output but observed with different geometry angle off-axis view. The SR is somehow amplified by the target magnetic (energy density $U_B$) of virtual photons target, while the ICS ones, in analogy, proportional to the target energy density radiation (both or self synchrotron one) $U_{rad}$. In synthesis
$P_{sync}\simeq\frac{4}{3}\sigma_\tau c\gamma_e^2U_B$, $P_{ICS}\simeq\frac{4}{3}\sigma_\tau c\gamma_e^2U_{rad}$, where $\sigma_\tau$ is the Thompson cross section, $\gamma_e$ the Lorentz relativistic factor of the electronic jet. In the presence of  a spread number distribution of the electron jet the same consequent scattered photons (ICS or synchrotron ones) will be opening their gamma jet within different cones and at different peak frequencies.

\subsection{The inner GRB jet opening angle shape}

There are therefore at least two (we shall not discuss here the eventual radial bremsstrahlung) different relativistic processes able to build a gamma jet: Inverse Compton Scattering, ICS and  Synchrotron Radiation, SR. It is very possible that both play a role in GRB jets. We shall consider here more in detail the SR motivated by the few well observed polarization of some gamma photons, as it has been recently proved \cite{2014Natur.509..201W, 2017Natur.547..425T}.
Our main assumption, following most recent article \cite{2017Natur.547..425T}, is that the earlier $\gamma$ jet, closer to its central source (NS or BH main attractive center of the collapse system) it is powered by a relativistic spiral UHE electrons driven by strong magnetic fields , ejecting while in spiral a sharp synchrotron emission. Later on, at highest distances, the jet could be also powered by self synchrotron or by ICS radiation. Let us assume, as most authors did, that the ejection is all collimated along the main axis of the jet. In this scenario the main opening angle $\Delta\theta$ of the gamma jet might be associated with the relativistic electron Lorentz factor
\begin{equation}
\gamma_e=1.957\cdot10^4\left(\frac{E}{10\,\mathrm{GeV}}\right)
\qquad\Delta\theta=\frac{2}{\gamma_e}=5.8\cdot10^{-3}\left(\frac{E}{10\,\mathrm{GeV}}\right)^{-1}
\end{equation}
This opening angle guaranteed a characteristic maximal beaming solid angle. Indeed this huge solid angle ratio $\frac{\Omega}{\Delta\Omega}\simeq4\cdot10^8\left(\frac{E}{10\,\mathrm{GeV}}\right)^2$, embrace the widest (see figure 4) apparent isotropic energy power spread of nearest (as the GRB170817A) and most distant GRBs at cosmic edges: a different alignment of the jet beam might explain with the same jet output the huge diversity of peak GRB luminosity within a unique model, the jet being powered by a NS tidally disrupt and absorbed by its binary companion (NS or BH).

Moreover, the same maximal Lorentz factor guarantee a maximal photon energy ($\sim$ GeVs) of the order of magnitude observed in rare hardest GRBs; for ICS we expect
\begin{equation}
h\nu_{\mathrm{GRB}}^{\mathrm{ICS}}\simeq\gamma^2kT_{\mathrm{GRB}}\approx6\,\mathrm{GeV}\,\left(\frac{E}{10\,\mathrm{GeV}}\right)^{2}\left(\frac{kT_\gamma}{15\,\mathrm{eV}}\right)
\end{equation}
while for the SR we find (assuming both extreme $\gamma$ hardness and softness):
\begin{equation}
h\nu_{\mathrm{GRB}}^{\mathrm{Synch}}\approx6.65\cdot10^{9}\,\mathrm{eV}\,\left(\frac{E}{10\,\mathrm{GeV}}\right)^{2}\left(\frac{B}{10^9\,\mathrm{G}}\right)
\end{equation}
\begin{equation}
h\nu_{\mathrm{GRB}}^{\mathrm{Synch}}\approx6.65\cdot10^{4}\,\mathrm{eV}\,\left(\frac{E}{30\,\mathrm{MeV}}\right)^{2}\left(\frac{B}{10^9\,\mathrm{G}}\right)
\end{equation}
In particular in the latter case the assumed high magnetic field ($B\sim10^5$ T) implies a very nearby surface location at NS-BH surfaces for the ejected beam. Naturally, as we already mentioned, we may imagine that the electrons follow a power spectrum $\frac{\mathrm{d}N_e}{\mathrm{d}E_edt}\propto E^{-\alpha}$, $\alpha\simeq2$ in an equipartition energy case, leading also to a rich number of lower energetic photons naturally much less collimated and at much weak luminosity.
This spectra will lead on average to a synchrotron radiation photon spectrum
$\frac{\mathrm{d}N_\gamma}{\mathrm{d}E_\gamma dt}\propto E_\gamma^{\left(\frac{-\alpha+1}{2}\right)}$

The whole effect allow us to have a spread conical jet at low soft gamma region respect to a more collimated jet at highest energy photons. The advantage is to have a more probable blazing by an off axis weak GRB as for the GRB980425 \cite{1999A&AS..138..507F} weak nearest event, respect a more collimated and brighter hard beam flash confined in most vast cosmic distances and volumes .

\section{The inner electron spiral  jet}
From all above arguments one will be tempted to conclude that the most wider and soft electron should be (for the photon component at $6\div8\cdot10^4$ eV comparable with the Fermi GRB170817A) within a solid angle $\frac{\Delta\Omega}{\Omega}\propto\gamma_e^{-2}\simeq(1/60)^2\simeq2.7\cdot20^{-4}$.
This wide cone makes reasonable an easier nearby GRB detection respect thinner cosmic ones $\frac{\Delta\Omega}{\Omega}\propto10^{-8}$, but its value is still very low and un-probable to be observed at the first NS-NS discover. If we assume that the very first NS-NS detection was observed already in a rare parameter edge (as shown in previous figures), the requirements that the solid angle will be pointing to us at $2.7\cdot10^{-4}$ probability rate is not acceptable. Naturally one may relax the electron energy component to a quite small values as, for instance a few $3$ MeV, obtaining a much wider spread angle; however the jet solid opening angle and the consequent probability to hit the Fermi telescope  it is still
of a few percent.

\begin{figure}[h]
  \centering
   \includegraphics[width=133mm,height=82mm]{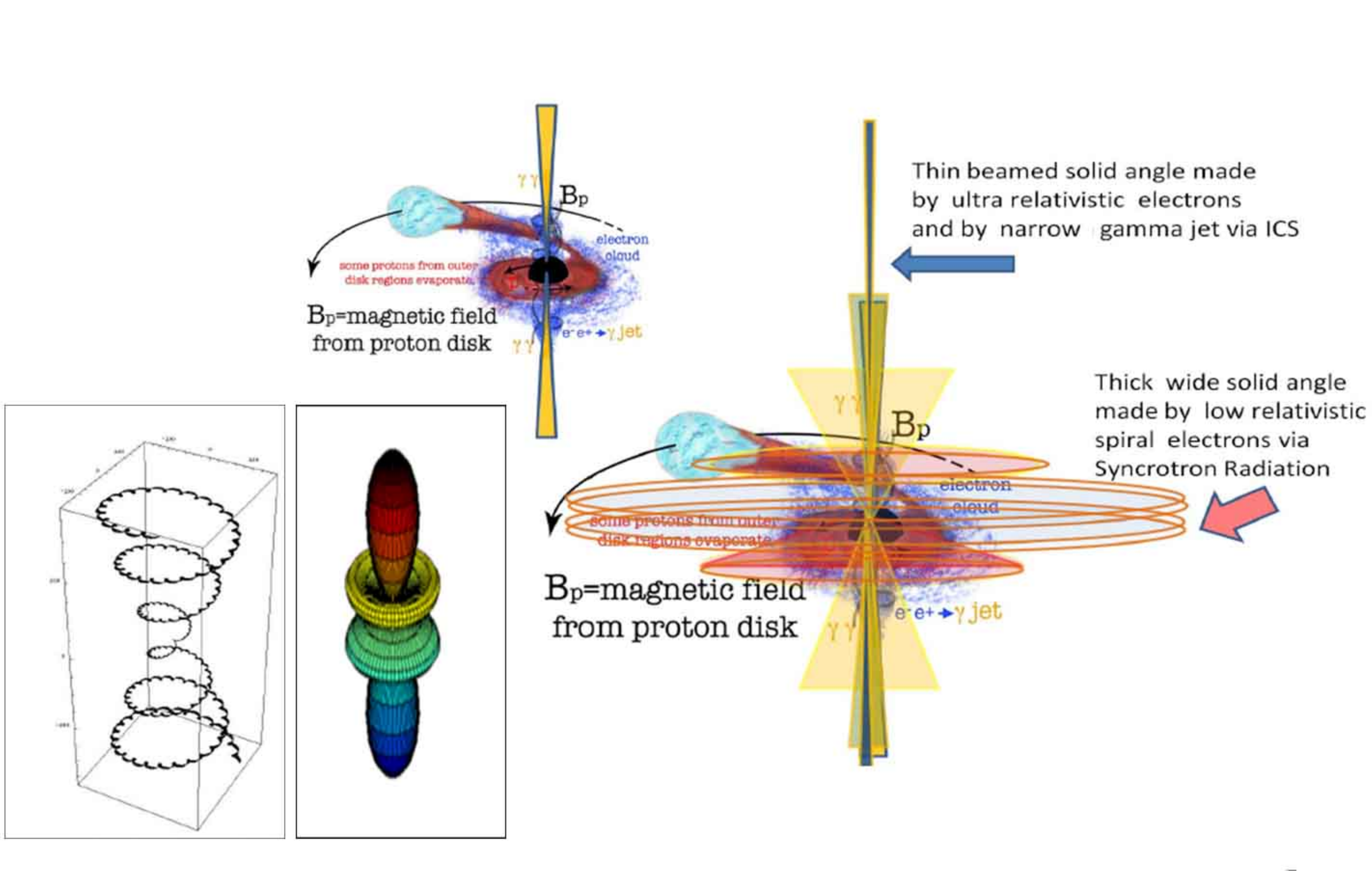}
   \caption{A simple schematic figure that summarize the NS NS collapse, fed by NS tidal fragment and powered at the main collapse into a beamed jet \cite{Fargion:2016}.
   The inner structure of the jet may also combine a circular spiral relativistic jet (ruled by Synchrotron radiation)  and the more collinear ultra-relativistic aligned electron jet shining a narrow hard gamma collinear jet. This hard gamma component may be ruled also by ICS or by self Synchrotron ICS. The persistent (but decaying) precessing twin jet may shine to us several times by its geometry shown in the lateral figure. The additional widest and weakest twin disk gamma cones offer a wider solid angle and a higher probability to shine respect inner beam jet as it may be the case for GRB 170817A.  }
  \label{Fig:ghirlanda}
\end{figure}

\subsection{The wide spiral twin cone \protect{$\gamma$} disks jet}
However there is a very powerful geometric novelty that is based on the very inner initial spiral electron radiation associated with the huge magnetic fields at NS surface and the probable prompt synchrotron radiation nature: the almost circular electron spinning (at the initial NS or BH base jet) makes they illuminating $\gamma$ along an almost planar disk whose solid angle spread not inside the same thin jet, but outside along a vast plane in a much spread  (twin, up and down) circular $\gamma$ solid angle: therefore the emission is not constrained into
$\frac{\Delta\Omega}{\Omega}\propto\gamma_e^{-2}$ but a much larger $\frac{\Delta\Omega}{\Omega}\propto2\pi/\gamma_e\sim1/60\left(\frac{E}{30\,\mathrm{MeV}}\right)^{-1}$.
 In this view the probability P to be observed is amplified into  P $\sim \frac{1}{60}(\frac{E}{30\,\mathrm{MeV}})^{-1}$ while the apparent luminosity is partially increased (with respect to the spherical isotropic case) only by nearly 60 times. Because the image it is specular (up and down twin jet) the probability to be observed it is twice as large: P $\sim\frac{1}{30}(\frac{E}{30\,\mathrm{MeV}})^{-1}$; a softer electron and a higher magnetic field, the GRB X component, as the afterglow probably observed (yet unpublished) by Chandra  satellite may reach higher value and visibility. Therefore the probability to be observed by Fermi as an off axis blazing event it is no longer a rare case, but it is within several percent or $10\%$ range.
 The time evolution of the electron jet spirals at larger distances (from its initial birth place where the B field B it is large and  electron energy is small) toward a high altitude and energy, but at lower magnetic field, leads to spiral electron rings  more and more collinear along the main jet, forcing the wide gamma twin disk emission toward a late more narrow beamed cone  $\gamma$ whose extreme solid angle may finally converge to $\frac{\Delta\Omega}{\Omega}\propto\gamma_e^{-2}$: the different Lorentz factor of the electrons $\gamma_e$  in the spectra offers huge extension of apparent luminosity that contains nearby $L_{\mathrm{GRB}}\sim10^{46}$ $~erg~s\textsuperscript{-1}$ observed off axis toward luminosity up to the most beamed (far cosmic ones) as large as $L_{\mathrm{GRB}}\sim10^{53}$ $~erg~s\textsuperscript{-1}$.

\section{Conclusions: the GRB170817A surprisingly an off axis gamma disk blaze?}

The need of an explanation for the eventual GRB170817A connection with GW170818 it is obvious and urgent.
We have shown the necessary high rate $\simeq 10^{4}$ $y^{-1}$ $Gpc^{-1}$ of such NS-NS merging to be consistent with the nearby location. These large values are not prohibitive. Also several bounds on chemical r-processes pollution might be accommodated  with the lucky GRB-GW connection. However the usual expected GRB beamed nature, even in the soft GRB  off axis view, seem implausible. Any  ad hoc jet opening angle for this event it seems to ignore the need of a consistent model also for a vast sample of apparent GRB Luminosity along far  Universe.
We did already considered for a similar rare low luminous and nearest GRB980425 event as an off axis
nature \cite{1999A&AS..138..507F}. However at the present the statistical rarety force to consider a more wide open solid angle of the early GRB burst  in order to make this lucky  GW and GRB connection an acceptable one.
  We have shown that in the view of a characteristic polarized GRB nature \cite{2017Natur.547..425T},
the Synchrotron Radiation, SR, may be better explained assuming a spiral electron (up and down) jet whose earliest shining photons are spread along a twin (up and down) mirror circular disk $\gamma$ cones. The consequent solid angle $ \frac{\Delta \Omega}{\Omega}$ of the earliest GRB170817A it is observable by a much wider than usual expected $\frac{1}{{\gamma}^{2}}$ and it is almost $\frac{1}{{\gamma}}$. The increase of the spread solid angle (and of the consequent probability to be observed) it is also increased by two or more order of magnitude respect a narrow beam jet. If the critical peak photons observed at GRB 170817A are formed by  electron at Lorentz factor $\simeq 6$

\begin{equation}
h\nu_{\mathrm{GRB}}^{\mathrm{Synch}}\approx6.65\cdot10^{4}\,\mathrm{eV}\,\left(\frac{E}{3\,\mathrm{MeV}}\right)^{2}\left(\frac{B}{10^{11}\,\mathrm{G}}\right)
\end{equation}

than the spiral electrons will shine in a wide twin disk and gamma ring where the observable solid
angle it is very wide, respectively  assuming a 3 MeV electron, $\simeq 30\%$ in its probability range. This enhanced off axis view it does not avoid the more energetic electron emission
at a more advanced stage within  a forward more collinear jet beam ; that component it is , as usual, mostly contained in a narrow solid angle as small as $\frac{1}{{\gamma}^{2}}$  leading to the huge amplified blaze, up to  hundred million times the $L_{GRB 170817A} \simeq 10^{44}$-$10^{46}$ $erg s^{-1}$ local almost non amplified  Luminosity, assuming ten or hundred GeV electron energy, as in above formulas for hard and powerful events.

In conclusion we learned from such a possible NS NS discover that nearly twice each minute in the Universe these merge occur.  These event rate may be compared with the SN explosions
(nearly several dozen each second). Their associated GRBs signals are often lost in the off axis geometry. Therefore only once a day the most beamed ones are revealed.
In the GRB 170817A as in GRB980425 the weakest tail of the twin jet disk $\gamma$ flash did reach our detectors. The OT and the multi precessing signature of their persistent $\gamma$ jet may give life to rarest hard  flashes in short time scale (mostly early $\gamma$ bursts) or sometimes to re-brightening and multi flares in late epochs in softer energy band (X, optical, Radio).

We imagine anyway that within the past years of records the Fermi (and other) $\gamma$ satellite detector there might be hidden  low luminous short GRB burst occurring off-axis in a very local Universe; these signals might be possibly been lost unnoticed because of the absence of any correlated GW coincidence. Future correlations between Fermi Chandra and Ligo-Virgo, may of course confirm and reinforce this new geometry interpretation of the puzzling GRB 170817A event. Naturally the whole merit for this great discovers  must be mainly addressed to the Ligo-Virgo as well to Fermi collaboration and their  commitment and sacrifices all along the last decades.
\section*{Acknowledgements}
The work by MK was supported by Russian Science Foundation and fulfilled in the framework of MEPhI
Academic Excellence Project (contract 02.a03.21.0005, 27.08.2013).

\label{lastpage}

\end{document}